**Numerical study of three-dimensional PIC for the surface plasmon excitation based on Drude model**

LIU La-Qun(刘腊群)[a], ZHAO Gui-Lian(赵桂莲), WANG Hui-Hui(王辉辉), LIU Da-Gang(刘大刚)

*School of Physical Electronics, University of Electronic Science and Technology of China, Chengdu 610054, People's Republic of China*

**ABSTRACT**: This paper explores the time-domain equations of noble metals, in which Drude model is adopted to describe the dielectric constant, to implement three-dimensional particle-in-cell (PIC) simulations for the surface plasmon excitation with the finite-difference time-domain method (FDTD). A three-dimensional model for an electron bunch movement near the metal film is constructed，and particle-in-cell (PIC) simulations are carried out with various metal films of different thicknesses. The frequency of surface plasmon obtained from PIC simulation is agreed with that from theory. Furthermore, the surface plasmon wave properties of excitation and propagation with the metal film is summarized by PIC results.

**PACS**：73.20.Mf, 78.67.-n, 78.20.-e.

# I INTRODUCTION

The surface plasmon wave, excited by electrons or waves at the noble metal surface[1], belongs to a surface electromagnetic wave. The surface plasmon excitation has broad application prospects[2] in the spectrum detection, biological/chemical sensing[3], light production, microscopic imaging, near-field optics[4], nano plasma

---

a) Electronic mail: 274626317@qq.com

lithography[5] and so on. Recently, the excitation methods of surface plasmon attract many interests, especially the excitation by moving electrons.

The surface plasmon excitation on the smooth metal surface cannot be excited by the light wave directly, because its vector is smaller than that of the surface plasma. Usually, some special structures such as prism coupling and grating structure are required to realize their wave vector matching. However, the wave vector of the moving electrons along the metal surface is larger than the light wave vector and it can match with the surface plasmon excitation's wave vector, so it has the ability to directly excite the surface plasmon excitation on the metal surface. In recent years, with the development of experimental techniques, both electronic vertical incidence metal surface and electronic movement along the metal surface have been studied[6-8]. Compared to other excitation methods, the projected field of moving electrons has a very wide spectrum and these electrons can excite the surface plasmon with quite wide frequencies on the metal surface by the vertical excitation. However, a coherent surface plasmon excitation can be stimulated by the parallel excitation. Besides, in the parallel excitation the frequency of the surface plasmon can be tuned easily by the adjustment of electron energy.

As a significant method to study the surface plasmon exciting by moving electrons, PIC simulations plays an irreplaceable role. In this paper, Drude model, commonly used in the study of the surface plasmon, is also used to describe the dielectric constant of the metal. And based on this dielectric constant, time domain equations are derived. These equations are applied to three-dimensional PIC simulations of the surface plasmon excited by electron beam parallel moving on the metal surface. On this basis, the law that the surface plasmons of the silver films of different thicknesses is excited by electrons in parallel is studied and it is well agreed with the theoretical calculation results.

The FDTD differential equations of the Drude model are derived in Sec. II. A three-dimensional PIC simulations model with an electron bunch moving in the direction parallel to the surface of silver film is introduced in Sec. III. Furthermore, in Sec. IV, the simulation results are analyzed and compared with the theoretical results. Finally, the summary is made in Sec. V.

# II DERIVATION OF THE FDTD DIFFERENTIAL EQUATIONS BASED ON DRUDE MODEL

In the Drude model, the dielectric constant $\varepsilon$ can be expressed as a functional form as follows[9],

$$\varepsilon(\omega) = \varepsilon_0 \left[ 1 - \frac{\omega_p^2}{\omega(\omega + i\gamma)} \right] \quad (1)$$

where $\omega_p$ is the plasma frequency, $\gamma$ is the average collision frequency, $\omega$ is angular frequency.

Since the dielectric constant is a variable related to the angular frequency in the Drude model, the original differential equation has not suited for the simulation and calculation of the material with a complex dielectric constant. The dispersion part of the medium parameters is needed to be reflected in the differential equation. From Eq.(1), the characteristic equation of the material in the Drude model can be expressed as,

$$\vec{D} = \varepsilon_0 \vec{E} + \vec{P} = \varepsilon(\omega)\vec{E} = \varepsilon_0 \left[1 - \frac{\omega_p^2}{\omega(\omega + i\gamma)}\right]\vec{E} \quad (2)$$

$$-\omega^2 \vec{P} - i\omega\gamma \vec{P} = \varepsilon_0 \omega_p^2 \vec{E} \quad (3)$$

where $\vec{P}$ is polarization. For $i\omega = -\frac{\partial}{\partial t}$, the time-domain form of equation Eq.(3) can be expressed as,

$$\frac{\partial^2 \vec{P}}{\partial^2 t} + \gamma \frac{\partial \vec{P}}{\partial t} = \varepsilon_0 \omega_p^2 \vec{E} \quad (4)$$

Introducing the polarization current density $\vec{J} = \frac{\partial \vec{P}}{\partial t}$, we get,

$$\frac{\partial \vec{J}}{\partial t} + \gamma \vec{J} = \varepsilon_0 \omega_p^2 \vec{E} \quad (5)$$

Applying difference operation to the Eq.(5) with Yee grid difference algorithm, we obtain the differential equation of the polarization current density $J_x$ in the x-direction,

$$J_x^{n+\frac{1}{2}}(i+\frac{1}{2},j,k) = \frac{2\Delta t \varepsilon_0 \omega_p^2 E_x^{n-1}(i+\frac{1}{2},j,k)}{2+\gamma\Delta t} + \frac{2-\gamma\Delta t}{2+\gamma\Delta t} J_x^{n-\frac{1}{2}}(i+\frac{1}{2},j,k) \quad (6)$$

In this paper, $i$, $j$, $k$ in the Yee grid difference algorithm are the labels of rectangular difference grid nodes in the direction of $x$, $y$, $z$, respectively. $n$ is the number of time steps of the current calculation.

In Maxwell's equations, the curl equations of the magnetic field can be expressed as,

$$\nabla \times \vec{H} = \varepsilon(\omega)\frac{\partial \vec{E}}{\partial t} = \varepsilon_0 \frac{\partial \vec{E}}{\partial t} + \frac{\partial \vec{P}}{\partial t} = \varepsilon_0 \frac{\partial \vec{E}}{\partial t} + \vec{J} \quad (7)$$

Applying difference operation to the Eq. (7) with Yee grid difference algorithm as well, we obtain the differential equation of the electric field $E_x$,

$$E_x^{n+1}(i+\frac{1}{2},j,k) = E_x^n(i+\frac{1}{2},j,k)$$
$$+\frac{\Delta t}{\varepsilon_0}\left\{\frac{H_z^{n+\frac{1}{2}}(i+\frac{1}{2},j+\frac{1}{2},k)-H_z^{n+\frac{1}{2}}(i+\frac{1}{2},j-\frac{1}{2},k)}{\Delta y}\atop\frac{H_y^{n+\frac{1}{2}}(i+\frac{1}{2},j,k+\frac{1}{2})-H_y^{n+\frac{1}{2}}(i+\frac{1}{2},j,k-\frac{1}{2})}{\Delta z}\right\}-\frac{\Delta t}{\varepsilon_0}J_x^{n+\frac{1}{2}}(i+\frac{1}{2},j,k)\quad(8)$$

Similarly, the differential equations of $J_y$, $J_z$ and $E_y$, $E_z$ can be derived and the differential equations of magnetic fields correspond with the solution in vacuum.

During the simulation, the polarization current density can be solved from Eq.(6) first, then it is substituted into Eq. (8) to solve the electric field. Since $\omega_p$ in vacuum is zero, its polarization current density equals to zero as well. However, the calculation of the magnetic field throughout the simulation as well as the interaction between electrons and electromagnetic fields still use the differential equations of Yee grid difference algorithm.

## III PARTICLE-IN-CELL SIMULATION MODEL

A model as Fig.1 shows that a pulsed electron bunch whose kinetic energy is 100keV and current density is 100kA/cm². Its thickness is 20nm in x-direction and the duration time is 0.2 femtosecond. The electron bunch at the top of the silver film moves along the positive direction of z-axis and it is parallel to the surface of the silver film, with a distance of 20nm from the thin film surface. Taking $\omega_p=1.2\times10^{16}rad/s$, $\gamma=1.45\times10^{13}Hz$ in the simulation, using Berenger perfectly matched layer (PML) as an absorbing boundary surround the x-direction and z-direction of the entire model. In the y-direction, cycle symmetry boundary is used to realize the simulation of semi-infinite region.

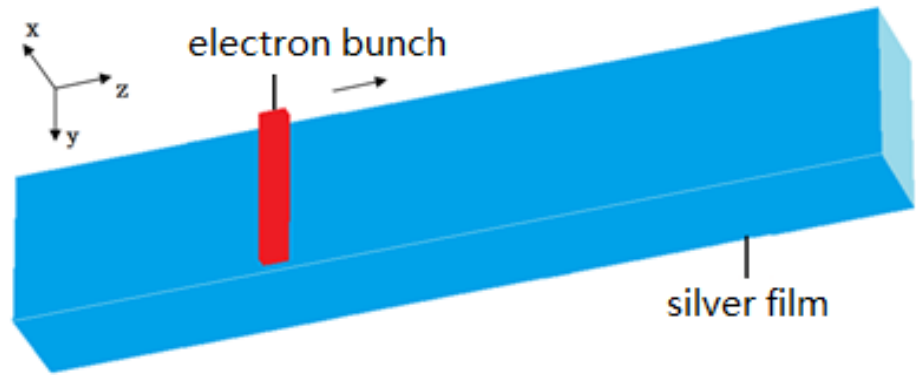

Fig.1 schematic model of three-dimensional simulation

# IV COMPARISON OF PARTICLE-IN-CELL SIMULATION RESULTS MITL THEOY

Firstly, we make a theoretical analysis of the surface plasmon excited by parallel electrons along the silver film before the particle-in-cell simulation results are given. According to the dispersion equations[10], Fig.2 shows dispersion curves of the silver thin film of different thicknesses (d=20nm, d=50nm, d=150nm). It shows the silver film of each thickness corresponds to two dispersion curves which are asymmetric mode and symmetric mode, respectively. With the increasing of the silver film thickness, the asymmetric mode and symmetric mode are increasingly near and the two dispersion curves are nearly coincident when the silver thin film having a thickness of 150nm. This is because the coupling strength of the surface plasmon between the upper and lower interfaces of the film is strong when the thickness of the film is thin. Then the two dispersion curves are more separate. However, the coupling strength is weak when the film is thick and the two dispersion curves are nearly coincident. In the figure, the solid line of larger slope represent the speed of light and the solid line of smaller slope is the electron line with 100keV. The intersections of the electron line and the dispersion curves of the silver films with various thicknesses are the frequencies of the stimulated surface plasmon when the electrons with 100keV moving in a direction parallel to the silver film.

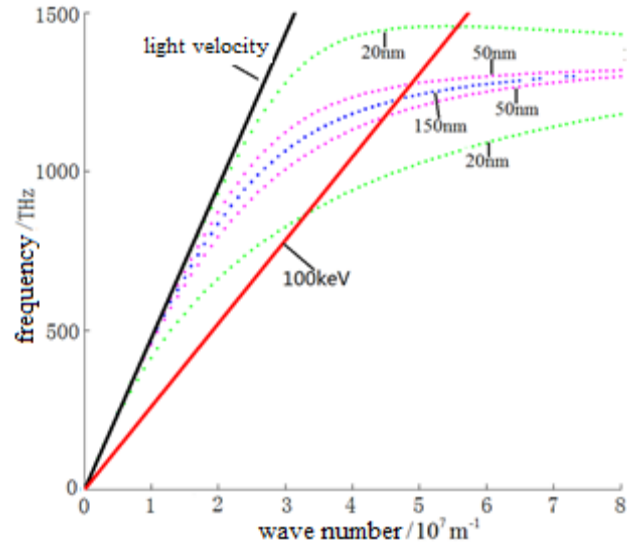

Fig.2    schematic of the dispersion relation of the silver film

Fig.3 shows the allelic figure of $E_z$ on the silver film surface at 15fs in the simulation when the thickness of silver film is 50nm. Fig.4 shows the distribution of $E_z$ along z-direction on the silver film surface at the same time. The two pictures indicate that the surface plasmon stimulated by electrons has obvious dual-band characteristics. Fig.5 shows the distribution of $E_z$ along x-direction at t=11.5fs, z=1.5 $\mu m$. Fig.3 and Fig.5 indicate that the surface plasmon stimulated by electronics group has an exponential decay in the direction perpendicular to the surface of the film and it propagates only in the direction parallel to the silver film surface. All these phenomena are consistent with the characteristics of the surface plasmon.

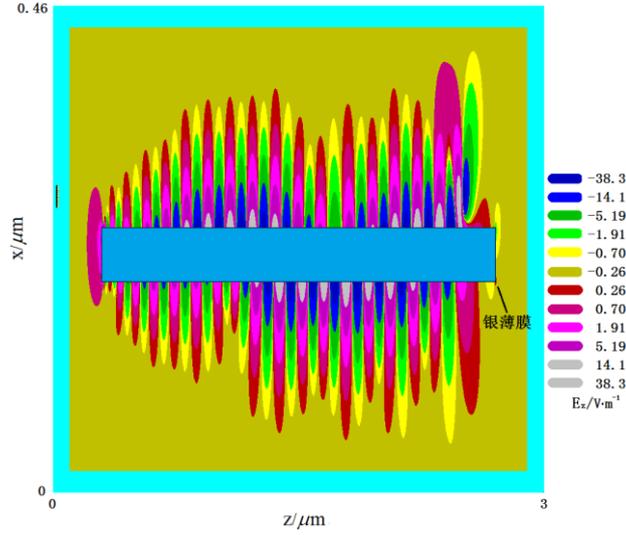

Fig.3 the allelic figure of $E_z$ on film surface at 15fs

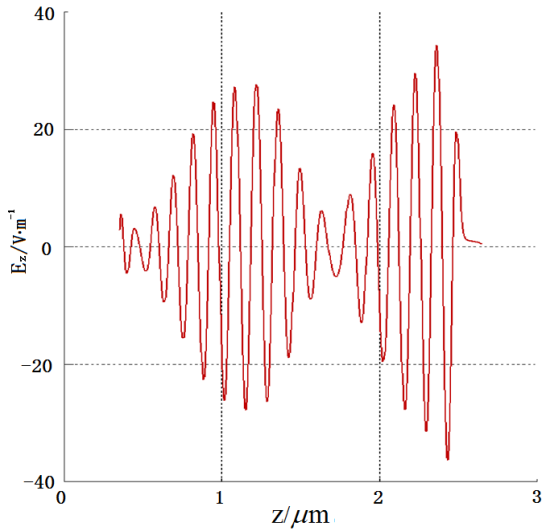 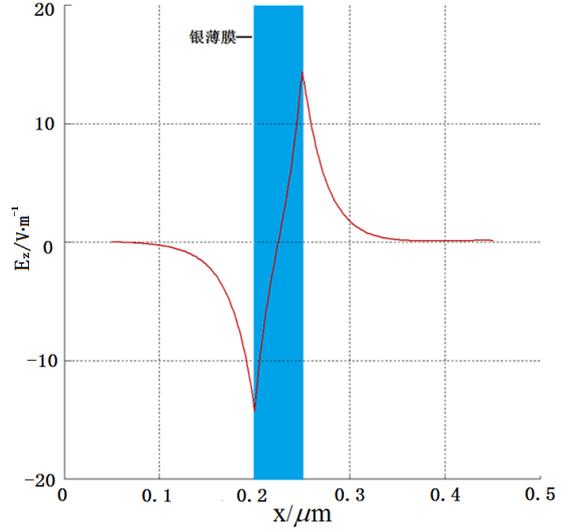

Fig.4 the distribution of $E_z$ along z-direction at 15fs

Fig.5 the distribution of $E_z$ along x-direction at 11.5fs

The time domain of $E_z$ at the upper surface of the film (z=1.5 $\mu m$) is shown in Figure 6. It can be easily seen that the amplitude of the electric field presents the characteristics of decay with time, which is the obviously characteristics of SPPs. Actually, the decay rate of electric field is related to the average collision frequency $\gamma$ in the dielectric constant and the larger the $\gamma$ is corresponding to the faster the attenuation and the shorter the duration of SPPs. Fig.7 is the spectrum of $E_z$ in the Fig.6.

It can be seen that the two modes of SPPs with frequencies of 1170THz and 1271THz are excited by electron bunch simultaneously.

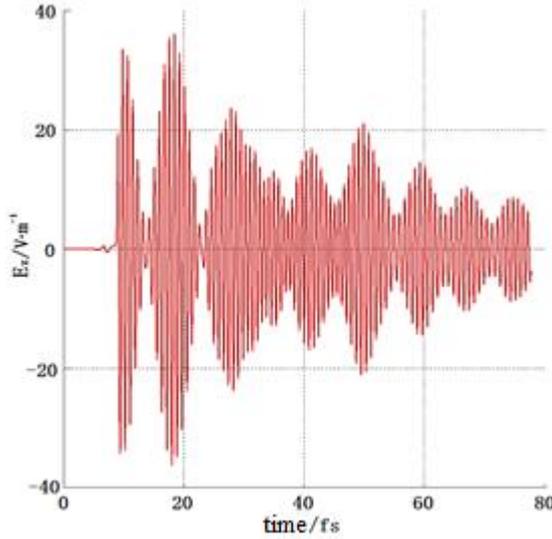 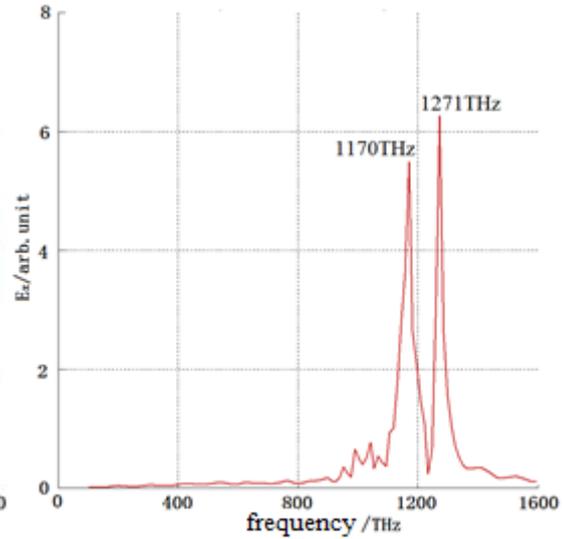

Fig.6 $E_z$ changes with time at the upper surface of the film

Fig.7 frequencies of $E_z$ at the upper surface of the film

Table 1 shows the numerical calculation results(the theoretical value) and the particle simulation results. The numerical calculation results is given by the intersections of the electron line and the dispersion curves of the silver films with various thicknesses. By contrast, it can be easily found that the variation trend of frequency with thickness is consistent with Fig.2. The two frequencies are close to each other with the increase of the thickness of the film and the particle simulation results are proved to be greatly accurate.

table 1 comparison between the theoretical value and the particle simulation results

| film thickness (nm) | theoretical value (THz) | | particle simulation results (THz) | |
|---|---|---|---|---|
| 20 | 862 | 1455 | 860 | 1455 |
| 30 | 1034 | 1363 | 1037 | 1365 |
| 40 | 1123 | 1307 | 1122 | 1308 |
| 50 | 1170 | 1273 | 1170 | 1271 |
| 60 | 1193 | 1253 | 1193 | 1254 |
| 70 | 1207 | 1241 | 1208 | 1244 |
| 80 | 1214 | 1234 | 1214 | 1233 |
| 150 | 1225 | | 1225 | |

# V CONCLUSION

The paper using the FDTD method achieves three-dimensional PIC simulations for the SPPs by deriving the time domain equations of dielectric constant in the Drude model. This method is used to simulate and research SPPs excited by electron bunch at the silver film surface. The simulation method is proved to be right by comparing the theoretical results with the particle-in-cell simulation results. However, researching on the surface plasmon with three-dimensional PIC simulations method faces a problem of quite slow computing speed which is caused by the large number of electrons and grids, and the short time step. Usually, a large-scale parallel computing is required to improve the computing speed.